\documentclass{article}

\usepackage[english]{babel}

\usepackage[letterpaper,top=2cm,bottom=2cm,left=3cm,right=3cm,marginparwidth=1.75cm]{geometry}

\usepackage{amsmath}
\usepackage{graphicx}
\usepackage[colorlinks=true, allcolors=blue]{hyperref}
\usepackage{tabularx,caption,ragged2e}
\usepackage{indentfirst}

\begin{document}
\setlength{\parindent}{8ex}

\begin{titlepage}
  \begin{center}
      \vspace*{1cm}
        \LARGE
      \textbf{Development and Validation of MicrobEx: an Open-Source Package for Microbiology Culture Concept Extraction.}

      \vspace{1.5cm}
      \normalsize
          
\textbf{Garrett Eickelberg$^1$}

\text{Department of Preventitive Medicine (Health \& Biomedical Informatics)}

\text{Feinberg School of Medicine,}

\text{750 N Lake Shore, Chicago, IL 60611, USA.}
\vspace{\baselineskip}

\textbf{Yuan Luo$^2$}

\text{Department of Preventitive Medicine (Health \& Biomedical Informatics)}

\text{Feinberg School of Medicine}

\text{750 N Lake Shore, Chicago, IL 60611, USA.}

\text{Electronic address: yuan.luo@northwestern.edu.}

\vspace{\baselineskip}
\textbf{L Nelson Sanchez-Pinto$^3$}

\text{Department of Preventitive Medicine (Health \& Biomedical Informatics)}

\text{Feinberg School of Medicine}

\text{750 N Lake Shore, Chicago, IL 60611, USA;}

\text{Department of Pediatrics (Critical Care)}

\text{225 E. Chicago Avenue, Chicago, IL 60611, USA.}

\text{Electronic address: lazaro.sanchez-pinto\@northwestern.edu.}

\text{Phone: 1.800.543.7362}

\vspace{2cm}

\vspace{0.8cm}

\textbf{Word Counts (excluding references, headings, tables, abstract, and figures):} 

2000

\vfill

 \end{center}
\end{titlepage}

\begin{abstract}
Microbiology culture reports contain critical information for important clinical and public health applications. However, microbiology reports often have complex, semi-structured, free-text data that present a barrier for secondary use. Here we present the development and validation of an open-source package designed to ingest free-text microbiology reports, determine whether the culture is positive, and return a list of SNOMED-CT mapped bacteria. Our rule-based natural language processing algorithm was developed using microbiology reports from two different electronic health record systems in a large healthcare organization, and then externally validated on the reports of two other institutions with manually-extracted results as benchmark. Our algorithm achieved F-1 scores $>$0.95 on all classification tasks across both validation sets. Our concept extraction Python package, MicrobEx, is designed to be reused and adapted to individual institutions as an upstream process for other clinical applications, such as machine learning studies, clinical decision support, and disease surveillance systems.

\end{abstract}
\vspace{0.25cm}

\section{Introduction}
Microbiology culture reports are relied upon for myriad healthcare applications ranging from guiding clinical treatment decisions to global disease surveillance. In a clinical setting, microbiology culture reports are helpful in answering if an infection is present and what organisms are driving that infection \cite{RN272}. Outside of the clinical setting, microbiology data are used to monitor disease outbreaks, improve healthcare operations (e.g. monitor nosocomial infection rates), and are leveraged in a variety of observational studies \cite{RN273, RN275, RN276, RN274}. Thus, the data within microbiology reports impacts clinical treatment and public policy decisions, and are therefore critical for secondary use \cite{RN277, RN273}. 

Unlike many other structured laboratory test results, microbiology culture reports are often complex, semi-structured reports that pose unique challenges for large-scale secondary use applications. Samples sent to a microbiology laboratory routinely undergo numerous tests, such as gram stains and antibiotic susceptibility tests, each of which have different turnaround times, can produce more than a single result, and need to be linked to the original accession number \cite{RN272, RN273}. Additionally, results from each test can include both quantitative and qualitative data, and need to be reported as they become available to facilitate treatment decisions \cite{RN272, RN273}. Unfortunately, although there are efforts to standardize reporting and analysis of clinical microbiology data, the suitability of existing microbiology reports for secondary use are hindered by reporting variability and analysis practices \cite{RN271, RN279, RN280}. Finally, microbiology reports contain varying amounts of protected health information as defined by the Health Insurance Portability and Accountability Act, thus limiting the flexibility of this data for data sharing projects. Therefore, there is critical need for informatic tools that can navigate microbiology report data challenges and extract information to facilitate their secondary use. The goal of this study was to develop, validate, and release an open-source microbiology concept extraction (MicrobEx) system to facilitate secondary use of microbiology reports.

\section{Materials \& Methods}
\subsection{Datasets}
The derivation datasets for this study were extracted from the Northwestern Medicine (NM) Enterprise Data Warehouse (EDW). The regular expressions and logic flow of our extraction system were developed using 216,372 raw free-text microbiology reports extracted from critical care patients treated at one of 10 Northwestern Medicine intensive care units between 1/1/2010-1/1/2020. To define microbiology reports, we queried the NMEDW and manually curated 235 unique procedures associated with microbiology culture orders. The collection of microbiology reports had highly heterogeneous formatting and lacked consistent template features such as concept-value pairs and table structures. Additionally, our corpora contained full microbiology reports, as well as individual microbiology components such as gram stains and antibiotic susceptibility reports. To address these challenges, rules were crafted to separate reports into sections wherever possible. For cultures with multiple report entries tied to the same accession number, only the notes with the latest report update time were selected for downstream processing and analysis. Testing and validation of our extraction system was performed on two external datasets with 119,789 expertly annotated free-text microbiology reports from University of Chicago (validation 1) and Ann \& Robert H. Lurie Children’s Hospital (validation 2). The validation sets of microbiological culture results were part of prior study and details have been previously published \cite{RN277}. The reports from both hospitals were annotated by the same senior clinical research coordinator. All four datasets included microbiologic cultures reports from blood, urine, respiratory, and cerebral spine fluid samples.

\subsection{Algorithm Overview}
\begin{figure}[htbp]
\centering
\includegraphics[width=0.95\textwidth]{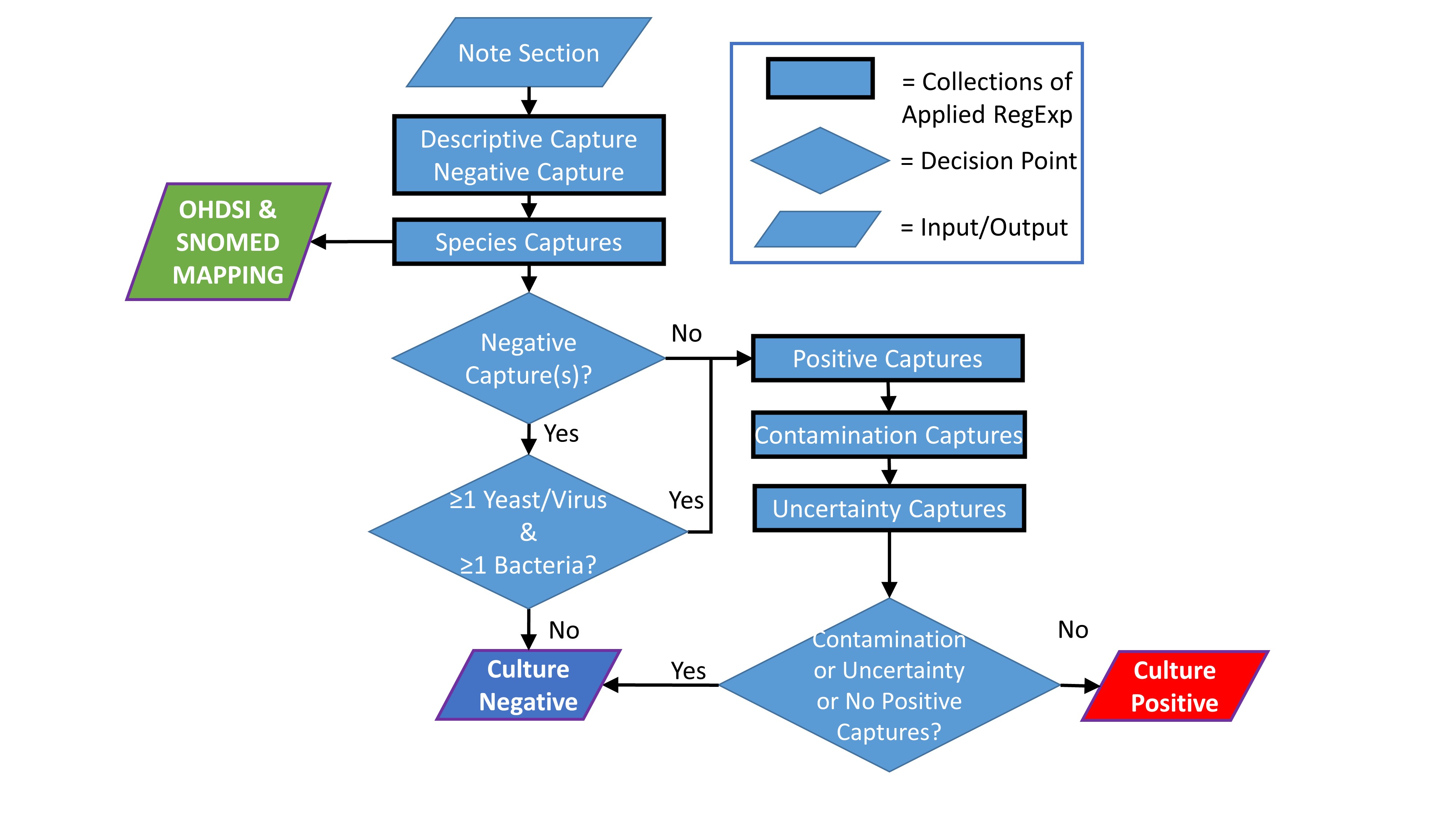}
\caption{\label{fig:fig1} The MicrobEx algorithm structure. The input of our algorithm is a whole or parsed section of a free-text microbiology culture reports. Within the algorithm, a series of regular expression collections are applied to the text input and the captures are associated with bacterial absence (negative), bacterial presence (positive), microbiological species, potential bacterial contamination, and uncertainty. Bacterial species captured are subsequently mapped to both OHDSI and SNOMED concept IDs. Hierarchical decisions are applied to the regular expression collection captures to categorize the culture as positive or negative for bacteria.}
\end{figure}

A summary of our algorithm workflow is presented in Figure \ref{fig:fig1}. Our concept extraction algorithm uses a comprehensive set of rules, as well as context, keyword, and morphologic features that capture overall bacterial infection status and identify bacterial species present in a microbiology report. Rulesets and regular expressions were developed through an iterative process based on document structural and context features in addition to clinical criteria and domain knowledge. For bacterial species captures, we wrote regular expressions to capture the genus and species for bacteria present in a dictionary of clinically relevant organisms collated from knowledgebases \cite{RN278, RN272}. Organisms captured were mapped to Observational Health Data Sciences and Informatics (OHDSI) and Systemized Nomenclature of Medicine (SNOMED) IDs via a dictionary included in the source code. The mapping dictionary for microorganism to OHDSI and SNOMED IDs was constructed by passing the collated microorganism list into Usagi software indexed on SNOMED vocabulary and restricted to class ‘ORGANISM’ and domain ‘OBSERVATION’\cite{RN282}. During each iteration, concept extraction performance was reviewed manually using a variety of different pattern occurrence-based audits on our training data sets. Customized regular expressions were created to capture remaining complex patterns. Each regular expression was developed with generalizability in mind to maximize dissemination and reusability. For all false positive and negative cases, we reviewed the associated case context, assigned a reason for misclassification. We addressed the cases by either refining existing rules or implementing new ones. This iteration process was repeated until all remaining uncaptured cases were caused by report noise, uncommon misspellings, or lack of report clarity \cite{RN264}.

\subsection{Validation}
Figure \ref{fig:fig2}. includes example reports annotated with extracted concepts, species, and estimated bacterial culture positive status. Both species extraction and binary bacteria positive culture status (yes/no) were evaluated as outcomes for validation of our algorithm and compared to the manually annotated results in the validation sets. For species extraction, we compared species captured across all report sections by our algorithm and the expert annotation. We encoded our binary outcome as positive if MicrobEx captured all the species identified by the expert. Similarly, positive bacteria culture status was assigned to all report sections and were compared to the expert annotation at the report level using a maximum function.

\subsection{Performance Benchmark}
In order to benchmark our algorithm’s performance against a well-established clinical natural language processing (NLP) tool, we applied MetaMap\cite{RN283} to both validation sets and built a rule-based decision workflow to predict positive bacterial culture status and capture bacterial species.

\subsection{Dataset customization}
To identify and address dataset-specific patterns capable of causing misclassifications, we audited our workflow as described in the \emph{use guide} prior to final validation. The generalizable regular expressions we added during the audits were both appended into the codebase prior to our validation studies.
The detailed code and Python package installation instructions have been made available at: https://github.com/geickelb/rbmce. See the \emph{use guide} section for Regular expression examples and a description on how to deploy and customize our package to a new dataset.

\section{Results}

\subsection{Validation}
Table \ref{tab:t1}. summarizes the distribution of positive bacterial culture status in the four datasets. The ratio of positive to negative cases across our training set predictions is consistent with that seen in the two curated validation sets.

\begin{table}[htbp]
\centering
\begin{tabular}{|l|l|l|}
\hline
{}             & Positive bacterial culture & Negative bacterial culture \\\hline
Derivation set 1 &  14,376 (20.7\%) &   55,065 (79.3\%) \\
Derivation set 2 &   23,549 (16\%) &    123,382 (84\%) \\
Validation set 1 &  2,185 (14.5\%) &  12,915 (85.5\%) \\
Validation set 2 &  7,391 (14.7\%) &   42,957 (85.3\%) \\
\hline
\end{tabular}
\caption{\label{tab:t1}Bacterial culture positive status distribution.}

\end{table}

Table \ref{tab:t2}. summarizes the validation results across both species and positive bacterial culture status classification tasks. The algorithm had excellent and consistent performance, with validation sets 1 and 2 having F1-scores of 0.99 and 0.96 for positive culture classification and species capture, respectively. To estimate the improvements made by introducing customized regular expressions from the data audits, each validation set was reanalyzed using a codebase with the associated regular expressions deactivated. From this, we estimate that culture positivity classification increased from 0.93 to 0.96 and 0.69 to 0.96 for validation sets 1 and 2, respectively. The addition of customized regular expressions was found to cause little-to-no effect on species capturing across both validation sets.

\begin{table}[htbp]
\resizebox{\columnwidth}{!}{\begin{tabular}{|l|l|l|l|l|r|r|r|r|}
\hline
{} & True Negative & False Positive & False Negative &  True Positive &  Precision &  Recall &    NPV &    F-1 \\
\hline

Validation set 1 & & & & & & & &\\
\rule{8pt}{0ex}
    Species capture &  12,463 (82.54\%) &      2 (0.01\%) &    209 (1.38\%) &  2,426 (16.07\%) &      0.998 &   0.921 &  0.984 &  0.958 \\
\rule{8pt}{0ex}
    Positive culture status &  12,909 (85.48\%) &      7 (0.05\%) &     22 (0.15\%) &  2,162 (14.32\%) &      0.995 &   0.990 &  0.998 & 0.992 \\

Validation set 2 & & & & & & & &\\
\rule{8pt}{0ex}
    Species capture &  42,391 (84.20\%) &      4 (0.01\%) &     68 (0.14\%) &  7,885 (15.66\%) &      0.999 &   0.991 &  0.999 &  0.995 \\
\rule{8pt}{0ex}
    Positive culture status &  42,950 (85.31\%) &      7 (0.01\%) &    606 (1.20\%) &  6,785 (13.48\%) &      0.998 &   0.918 &  0.986 &  0.956 \\
\hline
\end{tabular}}
\caption{\label{tab:t2}Infection classification and species capture performance across Validation sets.}
\end{table}

Supplemental Table 1. presents the results from our customized MetaMap based benchmarking algorithm against both validation sets. Across both positive culture classification and species capture, MicrobEx matched or surpassed the benchmark algorithm performance. These results suggest that our task-specific classifier can outperform more general-use clinical NLP tools like MetaMap.

\subsection{Error Analysis}
In the error analysis we identified a collection of five patterns in which our concept extraction workflow had the majority of errors. Figure \ref{fig:fig2}. presents annotated visual examples of the classification hierarchical logic for the different patterns observed, with examples for both correct classifications as well as misclassifications. Examples 5 and 6 depict the two most common types of false positive patterns and examples 7 and 8 present the most common patterns found in false negatives in the validation sets. We can summarize these patterns as a combination of multiple positive and negative organisms where the negative regex capture supersedes the positive captures, and the use of the term “contaminant” leading to a false negative classification.

\section{Discussion}
In this study, we developed and validated an open-source, rule-based framework to extract and map clinical concepts from microbiology reports to standardized terminologies to facilitate secondary use of microbiology reports. Our main finding is that our algorithm can reliably estimate binary bacterial culture status, extract bacterial species, and map these to SNOMED organism observations when applied to semi-structured, free-text microbiology reports from different institutions with relatively low customization.

\begin{figure}[htp]
\centering
\includegraphics[width=0.95\textwidth]{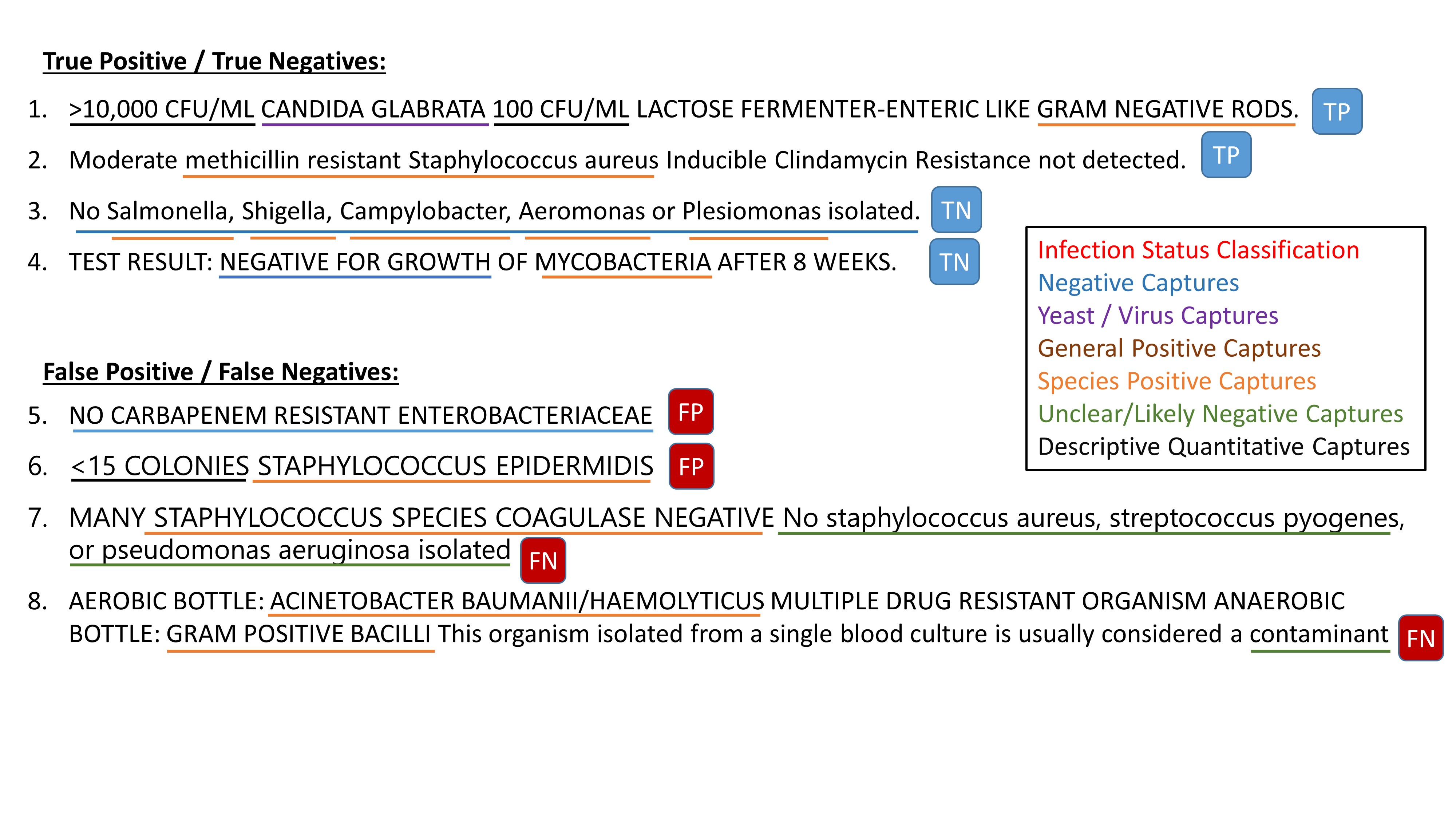}
\caption{\label{fig:fig2} Examples of annotated reports for validation and error analysis on validation set reports. Colored underlines correspond to parts of the report captured by the associated regular expression collection. For bacterial culture positive status classification, the concepts captured in each block are considered in a hierarchical decision structure according to Figure \ref{fig:fig1}. Examples 1 to 4 demonstrate algorithm annotation on cases found to be correctly classified as positive and negative. Examples 5 to 8 depict four examples representative of common misclassifications. Abbreviations: TP, true positive; TN, true negative; FP, false positive; FN, false negative.}
\end{figure}

Top performing rule-based concept extraction applications commonly employ a well-established clinical NLP tool that can map mentions to a corresponding medical concept(s) for broad medical corpora, such as cTAKES\cite{ RN284} and MetaMap\cite{RN283}. Like the well-established tools, MicrobEx performs concept matching by leveraging existing microbiology knowledgebases as described in Materials \& Methods. In contrast to these tools however, MicrobEx uses custom rules and regular expressions tailored to microbiology reports for dependency recognition and modifier detection. MicrobEx’s higher performance on bacterial positive culture status prediction suggests that for this classification task, MicrobEx’s more tailored approach provides advantages over an out-of-the-box approach using a well-established NLP tool. To further improve MicrobEx’s prediction performance, additional institution-specific customized rules could be added. Figure \ref{fig:fig2}. depicts four representative examples of cases misclassified for positive culture status that could be addressed with institution-specific custom rules.

To our best knowledge, three previously published studies have applied clinical concept extraction methods to microbiology notes  \cite{RN266, RN267, RN268}. Jones et al. \cite{RN267} applied a set of crafted rules to blood culture reports from the Salt Lake City Healthcare system to extract organism information, antibiotic susceptibilities, and infer if methicillin-resistant staphylococcus aureus (MRSA) was present. An evaluation was performed against approximately 10,000 expertly annotated reports to measure successful identification of MRSA. Matheny et al. and Yim et al. \cite{RN268, RN266} used hybrid and rule-based systems to capture combinations of microorganisms species and antibiotic susceptibilities from blood and multiple sample types, respectively. Our algorithm is notably different from the previously published systems in the following ways: (1) we estimate positive bacterial culture status, (2) our algorithm was designed to work with a variety of disparate microbiology report formats from different institutions, (3) we performed external validation on two expertly annotated microbiology datasets, and (4) our software is entirely open-source and available as a python package that can be further adapted to the reports of other institutions as described in the \emph{supplemental use guide} and supported by our results.

We recognize several limitations of our study. First, for users of this software, classifying positive culture status is the prediction task with the largest potential error. Compared to species extraction, which is largely string matching, estimating infection status requires significantly more complex logic. The hierarchical logic involved with positive bacterial culture status estimation is potentially susceptible to syntactic heterogeneity and report complexity, as depicted in Figure \ref{fig:fig2}. Additionally, we focused on bacterial cultures for the development and validation of the algorithm given the importance of antibiotic stewardship, antibiotic resistance, and bacterial sepsis in hospitalized patients. While our algorithm captures other microorganism species (including fungal and viral species), we did not validate the performance on those. Finally, we included logic to extract relevant quantitative and semi-quantitative concepts, however the performance of this was variable due to syntactic heterogeneity. As a result, we continue to provide quantitative captures as a feature of the MicrobEx algorithm, however these were not included in our validation.

\section{Conclusion}
In this article we detail the development, validation, and use of our open-source microbiology concept extractor (MicrobEx) algorithm and package. Our workflow achieved excellent performance in two independent validation sets with minimal customization. Our concept extraction Python package is designed to be reused and adapted to individual institutions as an upstream process for other clinical applications such as machine learning, clinical decision support, and disease surveillance systems.

\section{Acknowledgements}
LNSP and YL are co-corresponding authors. This research is partly supported by grants U01TR003528 \& R01LM013337 from the National Institutes of Health (Luo), grant 5T32LM012203\-04 from the National Library of Medicine (Eickelberg), and grant R01HD105939 from the National Institute of Child Health \& Human Development (Sanchez-Pinto).

\section{Competing Interest}
The authors declare that they have no known competing financial interests or personal relationships that could have appeared to influence the work reported in this paper.

\vspace{2.0cm}

\bibliographystyle{ieeetr}
\bibliography{rbmce_bib2}

\end{document}